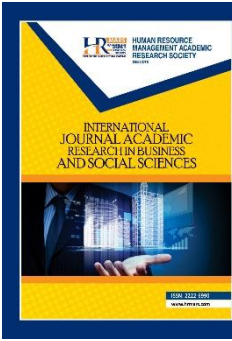
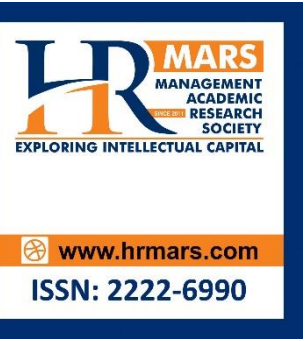

# Finding the Contextual Gap Towards Employee Engagement in Financial Sector: A Review Study

Habiba Akter, Ilham Sentosa, Sheikh Muhamad Hizam, Waqas Ahmed, Arifa Akter

**In-Text Citation:** (Akter et al., 2021)
**To Cite this Article:** Akter, H., Sentosa, I., Hizam, S. M., Ahmed, W., & Akter, A. (2021). Finding the Contextual Gap Towards Employee Engagement in Financial Sector: A Review Study. *International Journal of Academic Research in Business and Social Sciences*, *11*(5), 737–758.

# Finding the Contextual Gap Towards Employee Engagement in Financial Sector: A Review Study

## Habiba Akter, Ilham Sentosa, Sheikh Muhamad Hizam, Waqas Ahmed, Arifa Akter

UniKL Business School (UBIS), Universiti Kuala Lumpur, Kuala Lumpur, Malaysia.
Northern University Bangladesh, Dhaka, Bangladesh.
Email: habiba.akter@s.unikl.edu.my, ilham@unikl.edu.my, sheikhmhizam@unikl.edu.my,
waqas.ahmed@s.unikl.edu.my, miliarifa66@gmail.com

**Abstract**
This review paper identifies the core evidence of research on employee engagement, considering a stern challenge facing the financial sector nowadays. The study highlights the noteworthy knowledge gaps that will support human resource management practitioners to embed in the research towards sectoral context. Pertinent articles were selected through key search points. The Boolean logic (e.g., AND or OR) was applied to identify the relationship between search points and excerpt-related literature. The key search points covered the topic related to different terms of engagement for example "employee engagement" OR "work engagement" OR "job engagement" OR "organization engagement" OR "staff engagement" OR "personnel engagement" which were steered in diverse context particularly financial sector. Through critically reviewing the literature for the last 11 years i.e., 2009-2019, we discovered 91 empirical studies in financial sector. From these studies, we found the overall concept of engagement and its different determinants (e.g., organizational factors, individual factors, job factors) as well as its various outcomes (e.g., employee outcomes, organizational outcomes). We also formulated a conceptual model to expand the body of knowledge in the area of employee engagement for a better understanding of its predictors and outcomes. Besides, limitations of the study and future recommendations are also contemplated.

## Introduction
Workforces, notwithstanding the type of business, are deliberated as valuable assets for any organization. An optimistic, passionate and dedicated employee is the leading creative resource of an organization. Employees sometimes get involved themselves at job duties based on their knowledge along with emotive dedication, care, and obtainability in particular states. But companies always seek staffs who are usually eager to do their works regularly and who are very committed to their duties and responsibilities (Ahmed, Hizam, Akter, & Sentosa, 2020). According to Gruman and Saks (2011), engaged employees have been peddled as crucial to a company's achievement. Therefore, in these days of the competitive business market, the demands of organizational members have been gone beyond their salary where the exact core of employee engagement practices are more focused by the employers (Al-dalahmeh, Masa'deh, Khalaf, & Obeidat, 2018). A lot of researches provide evidence that the employee disengagement issue has been touted globally (Motyka, 2018). For example, in accordance with Gallup (2017), 85 percent of staff around the world are not actively engaged at all, whereas just 15 percent are entirely engaged at the workplace. Besides, a study by Sauerman (2019) stated that human resource (HR) policymakers nowadays are facing 12 challenges where employee engagement reached 41% which indicates that it is the highest momentum among all the challenges. A survey conducted by HRExchangeNetwork (2018) revealed  79% of employers opined that they have a high concentration to focus on the increasing engagement level of their employees.

In the bid of understanding and exploring the theme of overall engagement, various empirical evidence has been done on the topic related to employee engagement (Ghosh, Rai, & Sinha, 2014; Dajani, 2015; Aktar & Pangil, 2018; Monica & Kumar, 2018).  On the other side, a lot of systematic research reviews were done by previous researchers on engagement within the literature of business management. The reviews emphasize overall employee engagement (Omar, 2016); employee engagement and performance (Motyka, 2018); drivers of employee engagement and its effect on employee outcome (Bedarkar & Pandita, 2014); work engagement interventions (Knight, Patterson, & Dawson, 2016); work engagement in the context of Polish research (Pollak, Chrupała-Pniak, Rudnicka, & Paliga, 2017); a narrative synthesis of overall employee engagement (meaning, antecedents, and outcomes) (Bailey, Madden, Alfes, & Fletcher, 2017) and a critical review of employee engagement within the public sector (Fletcher, Bailey, Alfes, & Madden, 2019). There seems to have been none of these systematic review papers that paid attention to employee engagement, in general, especially the financial sector either as a combined or solitary notion.

On the other hand, prior researchers stated that human resource practitioners are doing constant research work on employee engagement for the requirement of business practice. Notwithstanding, there is a paucity of the constancy in definitions, measures, predictors, and consequences of employee engagement. Besides, there is little systematic review research on employee engagement to date in the worldwide context (Omar, 2016; Sun & Bunchapattanasakda, 2019). Jenkins and Delbridge (2013) first tried to scrutinize the impact of context on engagement.  The authors opined that the industrial sector and the condition of the marketplace affect human resource management (HRM) and the managerial function implemented to increase engagement level at the workplace. Writer (2017) underlined that employee engagement trends in the financial sector are identified as highly strong phenomena like other sectors. Furthermore, Borst, Kuyen, Lako, and de Vries (2019) suggested that more contextualization of engagement is required as if it is examined the engagement in a contextualized way, it can be better identified the influential factors and the explicit ways where engagement is practiced. This concern in terms of lacking contextualization as well as critical intuition has been leveled towards the research on engagement (Purcell, 2014; as cited in Fletcher et al., 2019), nevertheless, so far, no researcher attempts to critically review the scenario regarding the engagement while also taking into account the importance of context (Fletcher et al., 2019). The authors also provided evidence that there is the scantiness of systematic reviews regarding engagement within a sectoral context. Hence, this review paper focuses to synthesize the literature on employee engagement within the financial sector and concerning its several determinants and consequences.

**The Context of Financial Sector**
The financial sector is always acted as one of the most important sectors for the economic strength of any country. The importance of this sector lies as the "lifeblood" of financial action which plays a vital role in forecasting and executing economic strategy. To discourse the research gap, this review paper presents a systematic review based on prior empirical evidence regarding employee engagement in the context of the financial sector. In this research, the financial sector is meant the extensive sectoral area conquered by four sorts of financial organizations: a) banks, b) investment companies, c) insurance companies, and d) real estate firms. In this sense, the financial sector means a wide range of industries which is made up of financial organizations, brokers, and money markets which help to give the services to people to maintain their daily life.

A proper and efficient financial system is the pillar of a country's economy. When the system is operated properly, a country's economy can work efficiently without any difficulties. In this case, workforces are one of the most vital contributors to the achievement of the functional process of the financial system. But at present, employees' high turnover intention has become a challenging issue for the financial industry. So, it is an essential part for the employers in this sector to evolve. Besides, the financial industries require to readjust their HRM practices with the increasing engagement level of the workforces. If they fail to do that, they will fall into a risky situation due to the loss of millions of dollars in employees' turnover costs (Ufer, 2017). According to Writer (2017), financial organizations need to give priority to focusing on organizational goals through attracting and retaining talent. Because almost 75% of employees in financial sectors believe that employers can pay attention to them by giving them more opportunities to recover their services. For example, "Discover", the financial business industry is initiated in 1985 recognizing as the most familiar brand, taking as the highest position towards customer satisfaction with credit card companies which is knotted by J.D. Power in 2014. One of the main reasons for this achievement is that "Discover" prioritizes employee engagement first. On the other hand, almost 9000 working employees in the U.S.A concurred with 14000 working employees globally that engagement is a critical determinant of customer service, positive business outcomes, and retention (Kruse, 2015).

"Churn and Burn"- a common phrase is generally used to explain the high turnover rate in the industry (Ahmed, Hizam, & Sentosa, 2020). According to Ufer (2017), a recent survey indicates that this phrase is suitable to describe the financial business sector. The survey known as "Compdata" reveals that the financial sector has an 18.6% turnover rate that is one of the highest concerns among all sectors. Based on the PwC survey on millennials employed in financial business organizations, it was disclosed that just 10 percent of all millennials want to continue their present job for long periods. Besides, 42 percent of respondents opined they tried to look for new scopes where 48 percent were actively involved to find out other possibilities. The study further provided evidence that employees in financial organizations who left their current job in the year 2015 hold the positions just for 17 months. If it is compared to the number of 26 months in 2005 as well as 30 months in 1995, it is strongly evidenced that there is a high turnover rate in the financial sector. Blackburn, Way, and Auret (2020) explained that financial services will face challenges due to the forthcoming upsurge of disruption. Such as, it is the result of digitalization that the financial services have been facing more complexity regarding computerized procedures and roles as well as competitive business environment. To cope with this completive market, the financial business employers are struggling for attracting and retaining the skilled employees. Regarding this concern, they are endeavouring to motivate and engage their employees by giving more benefits to retain potential staffs. Hence, it is needed to evaluate the evidence regarding the relations between engagement and financial service consequences i.e., customer satisfaction, employee retention, organizational performance.

Unlike previous literature reviews, this research delivers evidence on how many empirical studies have been conducted on employee engagement within the financial sector to stimulate academic research in a different work setting on the interrelatedness of these notions. In addition, this systematic literature review covers the synthesis literature on overall employee engagement in the financial sector within diverse nations, thereby lengthening the literature sniff and finding gap acknowledgment extensive. These outcomes may notify the human resource practitioners and academics for developing interests, review and reform academic fields of research concerning these ideas. Finally, this research portrays a conceptual model that connects the concepts regarding employee engagement and provides a better comprehending of its different predictors as well as outcomes. Chhetri (2017b) opined that employee engagement is a concept of workforce behavior that requires precise search and necessitates a conceptual model for better understanding so that companies can ground their work system on it. Besides, the concept needs broadening in regarding relationship to its antecedents and consequences.

From the above discussion, it is cleared that this review paper studied comprehensive literature through a systematic review process to identify future research agenda regarding employee engagement in the context of the financial sector. This review paper looked for answering the following questions:

- How has the study contributed to comprehending the overall concept of employee engagement and its various predictors as well as outcomes in the context of the financial sector?
- What are the gaps of the study which exist in the current literature with precise reference to the financial sector?
- What are the possible guidelines for future researchers with meticulous reference to the employee engagement within the financial sector context that could be suggested?

**Methodology**

This paper is of a systematic review type that aims to identify the existing knowledge gap through delivering a structured analysis and the agglomerated outcomes. The overall literature search period was conducted from the year 2009 to 2019, in related databases, based on recent standards delineated in Moher, Liberati, Tetzlaff, Altman, and Group's (2009) guidelines for systematic review. Taylor & Francis, Emerald, Sage, Springer Link, Science Direct, ProQuest, EBSCO, Google Scholar, and Wiley Online were exploited as search engines for this review paper. The Boolean logic (e.g., AND or OR) was applied to identify relationships between search points and excerpt-related literature.

The key search points covered the topic related to different terms of engagement for example "employee engagement" OR "work engagement" OR "job engagement" OR "organization engagement" OR "staff engagement" OR "personnel engagement" which were conducted in diverse contexts particularly the financial sector. This research includes empirical studies written in English and peer-reviewed articles that examined the different predictors of employee engagement and its outcomes in the financial sector. The search strategy compiled 265 pieces of literature where 91 studies are identified as appropriate for this research criteria. Titles, outlines, key terms, introductions, findings, and discussion segments are studied for data accumulation on search points. Literature that has no covering on employee engagement in the financial sector context was excluded. Following, duplicate articles were removed, and the remaining articles were scrutinized for inclusion. Besides, conceptual papers, review papers and unpublished articles were excluded. The overall strategy for exclusion and inclusion criteria is outlined based on Moher et al.'s (2009) guidelines which are shown in figure 1 as follow:

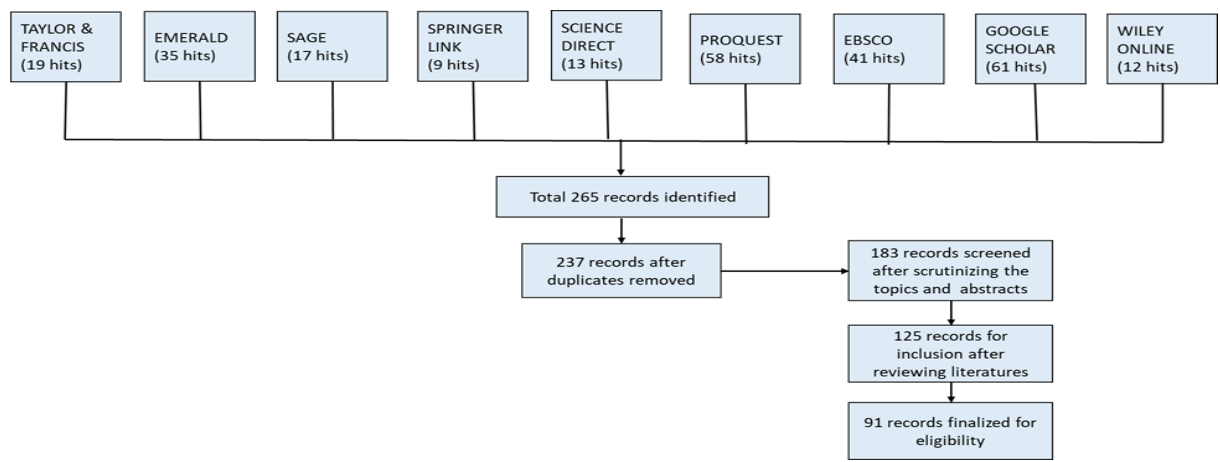

**Figure 1- Records Exclusion and Inclusion Criteria**

## Results and Discussions
### Year of Publication
This review recorded current empirical studies on employee engagement in the financial sector from the year 2009 to 2019. From the year 2009 to 2012 only 8 studies were included; however, 11 articles and 10 articles were chosen from the year 2013 and 2014 respectively. Only 23 research articles were recorded between 2015 and 2016, followed by 16 studies and 18 studies for 2017 and 2018 respectively. Lastly, only 5 literature for 2019 were selected. Figure 2 portrays the series of data recorded along with the number of publications identified from 2009- 2019.

**Country of Publication**
The authors further recorded the reviewed literature based on articles conducted across different nations in the context of employee engagement in the financial sector. The reviewed articles were conducted in 22 countries within 5 continents. This observation provides evidence of the widespread interest of numerous scholars in this area. Most of these studies have been conducted in Asia. Besides, 6.59% of studies have been done across different

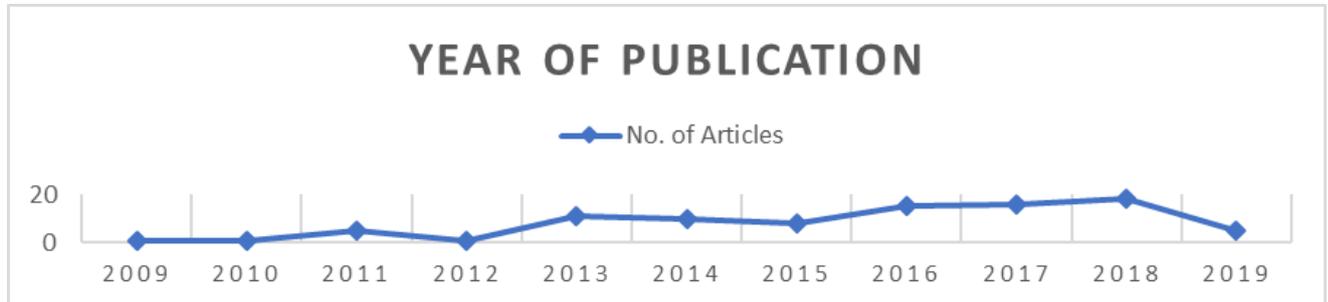

Figure 2- Publication Year of Reviewed Articles

European countries. The sum of articles in Africa accounted for 18.68% of all selected studies. The Middle East has also recorded 16.48% of total reviewed articles. Northern America and Oceania are shown low research regarding employee engagement in the financial sector. Figure 3 holds a country-based snapshot of overall reviewed articles.

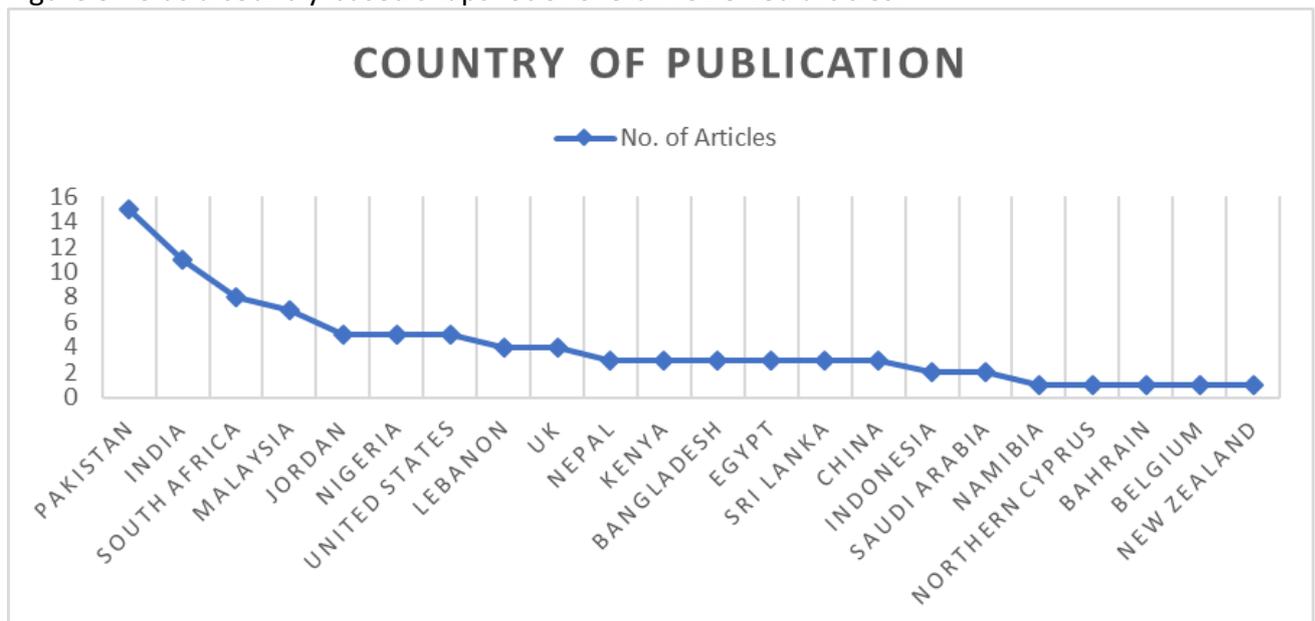

Figure 3 - Country of Reviewed Articles

**Concept of Employee Engagement**
Employee engagement is considered a wider notion than simply job engagement or work engagement in the existing literature. This paper discloses the different explanations used in the reviewed studies to define employee engagement in various ways that reflect different comprehending of staffs' engagement in every researcher's work. To conceptualize employee engagement, most of the reviewed articles used  Khan's (1990) concept, Schaufeli et al.'s (2002) concept, Saks's (2006) concept, Bakker's (2011) concept, Maslach and Leiter's (1997) concept, Hayes's (2002) concept, May et al.'s (2004) concept, and other scholars' concept. Based on their concepts of employee engagement, this review paper has clarified many enlightenments of overall employee engagement without applying any specific concept.

However, the different explanations of employee engagement defined by various scholars are displayed as follow in table 1:

**Table 1 - The Concept of Employee Engagement**

| Author(s) | Definition | Type of the Construct | Example of Research using the definition |
|---|---|---|---|
| Khan (1990); Rich (2006) | Employee engagement is the physical, cognitive and emotional state of workforces in which they have a connotation, confidence and security, physical and psychological abilities at the workplace regarding their work performance, safety, and availability. | | (Alvi & Abbasi, 2012), (Ghosh, Rai, Chauhan, Baranwal, & Srivastava, 2016) |
| Rich et al. (2010). | Employee engagement is referred to as a job-related construct i.e., job involvement and motivation at work. | | (He, Zhu, & Zheng, 2013) |
| Macey and Schneider (2008) | Employee engagement is the synthesis of employees' personality like attributes, emotional state, and positive behavior with situational facets like workplace environments. | | (Hassan & Ahmed, 2011) |
| Shuck and Wollard (2010); Shantz et al. (2013) | Employee engagement as a cognitive, emotive and behavioural condition. | Multidimensional phenomena. | (Rai, Ghosh, Chauhan, & Mehta, 2017) |
| Schaufeli et al. (2002) | Employee engagement as a positive, satisfying, job-related psychological state categorized by vigor (sense of high energy), dedication (sense of higher level of involvement), and absorption (sense of high concentration at work). | | (Karatepe & Aga, 2016) |
| Saks (2006) | Employee engagement as a different and exclusive concept comprising of cognitive, psychological, and behavioral constituents related to employee's work outcome. | | (Juhdi, Pa'wan, & Hansaram, 2013) |
| Rothbard (2001) | Engagement is a psychological state with the combination of dual crucial | Positive psychological state | (Aktar & Pangil, 2018) |
| | constituents including concentration and absorption. | | |
| **Nelson and Simmons (2003); Mauno, Kinnunen, and Ruokolainen (2007)** | Engagement as a positive psychological state of staffs at workplace where they realize their job to be directly meaningful, contemplate their job duties to be controllable and have confidence regarding upcoming workload. | | (Garg, Dar, & Mishra, 2018) |
| **Albrecht (2010)** | Employee engagement is a positive job-related emotional condition categorized by a real eagerness for contributing to the organizational outcome. | | (Ghosh et al., 2014) |
| **Baumruk (2004); Shaw (2005); Richman (2006)** | Engagement is the emotive dedication of the staff at the workplace. | | (Chaudhry, Jariko, Mushtaque, Mahesar, & Ghani, 2017) |
| **Christian, Garza, and Slaughter (2011); Alfes et al. (2012)** | Engagement is a comparatively continuing emotional state of the instantaneous willingness of individual involvement based on a person's skills or role performance. | | (Muduli, Verma, & Datta, 2016) |
| **Bakker (2011)** | Workforces who are highly energetic and involved in the work role and who believe themselves that they are capable to perform well according to their job requirements are identified as engaged employees. | | (Farid et al., 2019) |
| **Tritch (2003); Myrden and Kelloway (2015); Demirtas (2015)** | Employee engagement as workforces' enthusiasm, passion, and dedication towards their job role as well as workplace, the readiness to involve themselves along with extending their discretionary exertion for achieving organizational goals. | | (Ahmad & Gao, 2018) |
| **Maslach and Leiter (1997); Harter et al. (2002); May et al. (2004)** | The term "engagement" is identified as the state of employees' high energy, participation, and effectiveness which is the direct opposite of burnout dimensions namely, enervation, pessimism, and ineffectiveness; and both terms are dual trimmings of a continuum. | Opposite of burnout | (Banhwa, Chipunza, & Chamisa, 2014), (Shirin & Kleyn, 2017) |

**Type of Theories Applied**
Specifically, 64.84% of researchers applied theories to support their study. In total, 32.96% utilized a single theory (Lin et al., 2016; Aktar & Pangil, 2018); 25.27% applied two theories (Besieux, Baillien, Verbeke, & Euwema, 2015; Ghosh, Rai, Chauhan, et al., 2016; Bizri, 2018); and 6.59% combined three theories for supporting the research work (Rai et al., 2017). Most of the studies applied Social Exchange Theory (SET) and Job Demands–Resources (JD-R) model. These theories might be the most accepted theories regarding employee engagement. Because a stronger theoretical foundation has been found to clarify employee engagement within social exchange theory (SET) stated by Saks (2006) which explains the reciprocal relations between two parties (Presbit, 2017). Besides, Karatepe (2011) opined that the more hypothetical foundation for comprehending and scrutinizing employee engagement has been explained in the norms of SET. On the other hand, the JD-R model clarifies that employees are more likely to be engaged at the workplace if they have both job-related and personal resources. Because JD-R model differentiates between different resources (job-related or personal resources) and demands which can strongly foster the level of employee engagement (Bailey et al., 2017).

**Measurement Scales of Employee Engagement**
To measure the level of employees' engagement in the financial sector, different measurement tools are chosen by the prior academics, for instance, 53.85% of the reviewed articles used different versions of the Utrecht Work Engagement Scale (UWES) originated by (Schaufeli, Salanova, González-Romá, & Bakker, 2002). In its initial version, there are 29 statements including three dimensions of work engagement namely vigor, dedication, and absorption. "UWES-9" is the most commonly applied form chosen by the researchers (Rai et al., 2017; Garg et al., 2018). In addition, regarding employee engagement, 16.48% of the publications used Saks's (2006) scale (Biswas, Varma, & Ramaswami, 2013; Shah, Saeed, Yasir, Siddique, & Umar, 2017); 2.20% utilized Gallup's (2015) engagement scale (Banhwa et al., 2014); 4.40% of reviewed publications developed engagement scale (Haley, Mostert, & Els, 2013). Besides, 3.30%, 4.40%, and 3.30% used May, Gilson and Harter's (2004) tools (Mani, 2011), Rich et al.'s (2010) tools (He et al., 2013), Khan's (1990) engagement scales (Imam & Shafique, 2014) respectively. Furthermore, precisely 12.09% applied others' measurement tools of employee engagement such as Thomas's (2007) tools (Dajani, 2015); Lee's (2012) scale (Hassan, Hassan, & Shoaib, 2014); Fine et al.'s (2010) scale (Shaikh & Akaraborworn, 2017); Wellins, Bernthal and Phelps's (2004) scale (Muduli et al., 2016); Robinson et al.'s (2004) scale (Busse & Regenberg, 2018); Towers Watson's scale (2010) (Besieux et al., 2015).

**Data Analysis Techniques**
Precisely 92.30% of the reviewed studies dominated the quantitative method, 4.40% employed the qualitative method and 3.30% preferred mixed methods. The researchers of the reviewed studies, for testing their hypotheses, applied different techniques of data analysis. One of the most common techniques, structural equation modeling (SEM) including path analysis, exploratory factor analysis (EFA), confirmatory factor analysis (CFA), and partial least-squares analysis; used in 34.07% of reviewed articles (He et al., 2013; Els, Viljoen, Beer, & Brand-Labuschagne, 2016; Aktar & Pangil, 2018). 54.95% of the publications dominated Pearson correlation coefficient and Multiple Regression analysis (Ghosh et al., 2014; Thavakumar & Evangeline, 2016; Busse & Regenberg, 2018; Garg et al., 2018). Besides, only 3.30% of reviewed articles used other data analysis techniques namely t-Test and Bayesian methods (Monica & Kumar, 2018). Almost 38.46% of reviewed studies employed a single data analysis technique, 51.28% had two data analysis techniques and 10.26% dominated more than two techniques.

**The Predictors of Employee Engagement**
Eighty articles involved reference to the predictors of employee engagement. The reviewed findings disclosed that these can be categorized into six groups: psychological factors, job factors, interpersonal relations factors, individual factors, environmental factors, and organizational factors.

- *Psychological Factors:*

9.89 percent of reviewed studies experienced the association between psychological factors and employee engagement. For example, Malik and Khalid (2016) found a strong negative relation between psychological contract breach and work engagement (r = −0.76, p < 0.05) which specified that if staffs perceive psychological contract breach, it leads to lower levels of engagement at work. Alvi, Gondal, Ahmed, and Chaudhry's (2014) research exposed that employee empowerment was the strong forecaster of employees' job engagement. The study also confirmed that employee empowerment can cause a 34.4 % change in employees' job engagement. On the other hand, a positive correlation between psychological capital and employee engagement; and a negative relation between psychological contract breach and employee engagement was found in two types of research within financial sectoral context ; (Shirin & Kleyn, 2017; Asif, Khan, & Pasha, 2019).

- *Job Factors:*

4.40 percent of articles examined the association between job factors and employee engagement. Research done by Rai et al. (2017) evidenced that the association between job characteristics and work engagement was significant (p<0.01). Taipale, Selander, Anttila, and Natti (2011) did a study on work engagement in European countries' contexts where their results showed that demands decreased work engagement, while autonomy and support increased it. Hence, the study evidenced that a weak relationship existed between work engagement and work demands while work autonomy and social support strongly predicted work engagement.

- *Interpersonal Relations Factors:*

13.19% of reviewed articles studied the link between interpersonal relationships and employee engagement. Chaurasia and Shukla (2013) showed a positive relationship between the leader-member exchange relationship (LMX) and employee engagement. Further, a study done by Ghosh, Rai, Singh, and Ragini (2016) provided evidence that managerial support and co-worker support were significant determinants of employee engagement.

- *Environmental Factors:*

Precisely, seven studies (7.69%) tested the link between environmental factors and employee engagement; six studies found a positive link while the rest showed a negative relation of environmental factors to employee engagement such as a study concluded that working conditions, health, and safety positively and significantly influenced employee engagement (Banhwa et al., 2014). But, Mboga and Troiani (2018) found a negative relationship between work environment and employee engagement.

- ***Individual Factors:***

Thirteen articles explored the connection between individual factors and employee engagement. All these literature in each of the subsequent areas provided evidence of a positive relationship with employee engagement: work-life balance (Venkatesh & TA, 2014), person-organization fit (Chhetri, 2017b), self-directed learning, and employee education (Nadeem, Ghani, & Shah, 2017), self-consciousness (Rothmann & Rothmann, 2010), religiosity (Bakar, Cooke, & Muenjohn, 2016).

- ***Organizational Factors:***

Thirty-five articles explored the link between organizational factors and employee engagement. Some showed a strong positive relationship between different organizational factors and engagement. For instance, a recent study found leadership was the better predictor of employee engagement, where it explained 62.4% of the total variance of employee engagement. Rewards and recognition, work policies, and procedures appeared to be almost the same powerful predictor of employee engagement; where it explained 12.2% and 12.1% respectively of its total variance. Lastly, training and development appeared as the least predictive factor of employee engagement (Dajani, 2015). Besides, positive relations appeared between the subsequent factors and engagement in single articles: performance management, personal development opportunity, remuneration, career management, and organizational learning (Mokaya & Kipyegon, 2014).

A study done by Ghosh et al. (2016) showed rewards and recognition were a strong predictor of employee engagement. Muduli et al. (2016) evidenced that HPWS was strongly associated with employee engagement. Further, four studies showed a positive association between human resource management (HRM) practices and engagement (Aktar & Pangil, 2017, 2018). In addition, a positive link between different leadership styles and engagement was shown in seven articles such as transformational leadership (Naeem, Lashari, & Rana, 2017); ethical leadership (Ahmad & Gao, 2018); leadership behaviors (Xu & Thomas, 2011). A few pieces of literatures evidenced a weak relationship between some organizational factors and engagement. For example, a study conducted by Gummadi and Devi (2013) reflected that the relationship of training and development, as well as rewards, were not significant and weak leading to only a 1% possibility of an impact on employee engagement. On the other side, one article showed a negative association between organizational politics and engagement (Javed, Gulzar, & Hussain, 2015).

**Mediating Role of Employee Engagement**

35.90 percent of reviewed articles used employee engagement (job engagement, organization engagement) as a mediator. Precisely, 17.95 percent of articles used work engagement as a mediating variable. For example, a study conducted among banking employees showed that job engagement mediated the relations between perceived organizational support (POS) and task performance, POS and organizational citizenship behavior (OCB), POS and counterproductive work behavior (CWB), core self-evaluation (CSE), and task performance, CSE and OCB, CSE and CWB. Moreover, results revealed that the direct effect of CSE on task performance, CSE on OCB, CSE on CWB became significant when job engagement controlled such relationships, thus suggesting partial mediation. On other hand, the outcomes of the study also proved that the direct effect of POS on task performance, POS on OCB, POS on CWB became nonsignificant when job engagement controlled such relationships, thus suggesting full mediation (Chhetri, 2017a). Another empirical study revealed that work engagement fully mediated the effects of organization mission fulfillment (OMF) and POS on job performance (Karatepe & Aga, 2016).

On the other hand, only 3.85 percent of reviewed studies used organization engagement as a mediating variable. For instance, a study done by Juhdi et al. (2013) indicated that organization engagement partially mediated compensation, career management, person–job fit, performance appraisal, and job control to turnover intention. Moreover, 14.10 percent of studies identified that overall employee engagement mediated the relation between different antecedent variables and consequent variables. For instance, Ghosh et al. (2016) hypothesized in their study that employee engagement had a mediating effect between rewards and recognition and organizational commitment (normative commitment). Outcomes of the study provided evidence that the relationship between rewards and recognition and normative commitment had become smaller when employee engagement controlled such relation, which suggested a partial mediating role of employee engagement existed between rewards and recognition and normative commitment. Furthermore, Akhtar, Nawaz, Mahmood, and Shahid (2016) confirmed that employee engagement played a mediating role in the relation between high-performance work practices (HPWPs) and employee performance. Their findings also proved a significant effect of HPWPs (i.e., training, employee empowerment, rewards) on employee performance but the intensity of the effect has been narrowed in the presence of mediating role of employee engagement. So, it is ensured that employee engagement performed as a partial mediator in the effect of HPWPs on employee performance. The study conducted by Chaurasia and Shukla (2013) outlined that as a mediating mechanism, employee engagement (job engagement, organization engagement) linked LMX (leader-member exchange relationship) to work role performance. Lin et al.'s (2016) study evidenced that employee engagement had an indirect effect between future work self salience (FWSS) and two performance indicators i.e., supervisor-rated and sales performance.

**Contextual Variables/Moderators**

Out of the 91 reviewed studies, only 13.19% of articles used contextual variables or moderators. For instance, Shah, Said, and Mahar (2019) provided evidence organizational trust (agreeableness) and managerial support positively influenced the level of workforces' engagement in a workplace. Their results confirmed organizational trust plays a moderating role between perceived supervisor support and employee engagement (where R=0.527; P<0.05). Another study conducted in the Pakistani banking sector context empirically proved that power distance orientation (extraversion) moderated the impact of ethical leadership on job engagement through psychological empowerment. The study also showed that for low power distance orientation, the connection between ethical leadership and psychological empowerment is stronger than high power distance orientation (Ahmad & Gao, 2018). Imam and Shafique (2014) postulated job stress (neuroticism) performed as a moderator in the effect between employee engagement and employee outcomes (job satisfaction, organizational commitment) where their postulation is rejected.

**The Outcomes of Employee Engagement**

The authors determined different outcomes of employee engagement which can be categorized into two groups: employee outcomes and organizational outcomes.

- ***Employee Outcomes:***

Exactly, thirty-three (36.26%) articles explored the association between employee engagement and employee outcomes. Most of the reviewed articles disclosed a positive relationship between employee engagement and a variety of employee outcomes, such as job performance, organizational commitment, turnover intention, organizational citizenship behavior (OCB), and counterproductive work behavior (CWB). For instance, Karatepe and Aga, (2016) tested job performance as an outcome of work engagement. Their results exposed that work engagement positively affected job performance. The authors further demonstrated work engagement appeared as the most proximate inspirational factor to job performance. Banhwa et al. (2014) showed that the correlation between employee engagement and OCB was found to be significant. Dajani (2015) concluded that employee engagement appeared to be an important determinant for job performance, where it explained 14.9% of its total variance. Furthermore, Lin et al.'s (2016) study evidenced that employee engagement had a positive correlation on two performance indicators i.e., supervisor-rated and sales performance

Chhetri (2017a) did research among banking staff in the Nepalese context to determine the determinants and consequences of job engagement. Measurement of the engagement explicated there was a moderate relation between job engagement and task performance ($R^2$= 0.39, p< 0.01) and a moderate correlation existed between job engagement and OCB ($R^2$ = 0.41, p< 0.01) but there was a very poor relation between job engagement and CWB ($R^2$ = 0.24; p,0.01). Further, a study done by Mahesar, Chaudhry, Ansari, and Nisar (2016) provided evidence that job satisfaction, turnover intentions, and organizational commitment were found as positive and significant outcomes of employee engagement. For example, the correlation matrix revealed in the study that employee engagement is positively and significantly associated with job satisfaction (r =.10, p<.05), turnover intentions (r =.34, p<.01), and organizational commitment (r =.10, p<.05). Chaurasia and Shukla (2013) found in their study that variance ($R^2$ = 82%) in work role performance of employee engagement is very high that assured employee engagement played a vital role in the performance. Through Krishna and Murthy's study, employee performance has been found as an important consequence of employee engagement (Krishna & Murthy, 2015).

- ***Organizational Outcomes:***

Further, out of the 91 reviewed articles in the context of the financial sector, only 9.89% of articles found a positive link between employee engagement and organizational outcomes. For instance, Muduli et al (2016) conducted survey research among 600 Indian banking staff where results showed a positive link of employee engagement to organizational performance. Another research showed all the dimensions of employee engagement including vigor, absorption and dedication positively and significantly forecasted organizational performance. The research also explained that vigor had a high contribution in forecasting organizational performance, followed by absorption and then dedication (Al-dalahmeh et al., 2018). Zameer, Wang, Yasmeen, Mofrad, and Waheed (2018) did survey research among 522 responses (261 employees and 261 customers of the banking sector) in which they found employee engagement is the most powerful indicator that has a strongly positive effect on the corporate image and customer satisfaction.

Through reviewing existing literature on employee engagement in the financial sector, this systematic review depicted a conceptual model to expand the body of knowledge in the area of employee engagement and its different predictors as well as outcomes. The findings of the review lead to the subsequent model drawn in figure 4.

## Limitations

No research article is free from limitations (Hizam, Akter, Sentosa, & Ahmed, 2021). This review paper has also some limitations like other research papers. The limitations are related to the method used for choosing articles for scrutinization. The first one considers the criteria which had to be set up for the empirical research in the financial sector context to be eligible for the assessment (screened of articles based on rigorously explained information in their extracts, key terms, and findings). The second restraint is linked to the limited number of online search engines that asserted the search strategy. Lastly, this research paper does not contain conceptual papers, review papers, and unpublished works which possibly could boost it regarding theoretical facets of the notion of employees' engagement at work (Motyka, 2018).

## Directions for Future Research

This study, with the analysis of empirical research, sought to investigate how employee engagement in association with its antecedent variables and consequent variables has been studied in the financial sector. This research paper confirmed that, though employee engagement has been conducted widely in the financial sector, the results also find out noteworthy study gaps worth studying. For instance, in the case of research design, there is a high priority for choosing a quantitative method instead of a qualitative method. So, studies using a qualitative research design may be preferred by further researchers. In addition, though most research articles used UWES instruments for measuring engagement, there is still a paucity of measurement tools (Motyka, 2018). Hence, it is suggested to future studies for focusing more on the measurement scale of employee engagement because, at present, standard measurement instruments may have constraints.

On the other hand, a lot of studies drew their attention to the determinants and outcomes of employee engagement in the financial sector. Most of these studies strongly preferred organizational factors (e.g., rewards and recognition, management style, training and development, growth opportunities, decision-making systems) over individual factors (e.g., employee education, personal skills, and abilities). Therefore, it is suggested that further studies can focus on potential individual factors which will support boosting employee engagement like organizational factors. Further, as the mediating and/or moderating variable used among employee engagement and its predictors along with its outcomes by previous researchers has been found limited, future researchers can use a mediating and/or moderating variable in such relationships.

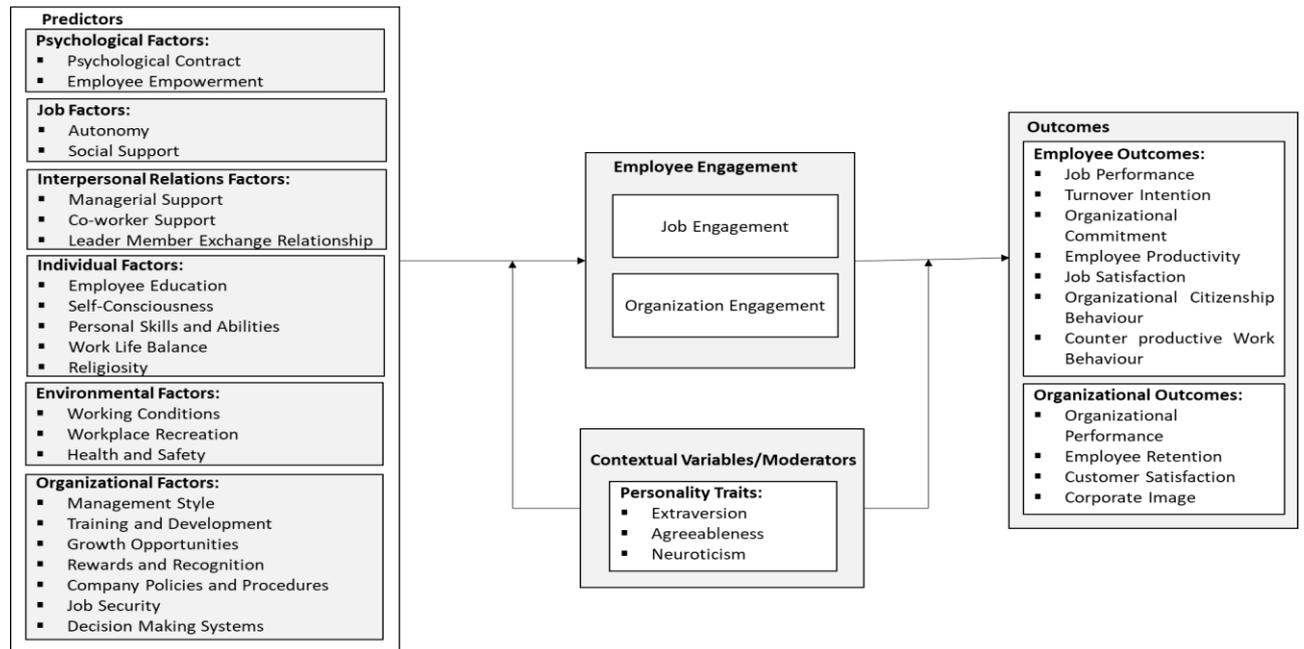

**Figure 4 - Employee Engagement Model Covered by its Predictors, Outcomes, and Contextual Variables.**

## Conclusion

This systematic review paper combines empirical articles on employee engagement which addresses issues of contextualization in terms of the financial sector. Although a wide range of empirical evidence has been found regarding relevant predictors and consequences of engagement, there seems to be a distinct dearth of attention to the precise contextual issues which can be essential in influencing the knowledge of engagement within the financial sector. However, this review has identified that there are a lot of studies have been found to discourse these issues. From systematically reviewing the scenario base, the authors find out the crucial questions along with further research agenda. Because, if it is not discoursed by HRM practitioners, it may lead to a deceptively simplistic understanding in terms of engagement and its effect on organizational goal as well as its beneficiaries. Our sequential reviews can help the comprehending of determinants and outcomes of employee engagement from wider and more diverse contexts; and aid to obtain insight into reality, constrictions, and solutions in increasing employees' engagement at work. Overall, this review paper confirms the growing attention of academics studying in the financial sector context regarding the topic of employee engagement.